\documentclass[sigconf]{acmart}
\usepackage{booktabs} 
\usepackage{amsmath,url,graphicx,amssymb}
\usepackage{amsfonts}
\usepackage{ctable}
\usepackage{multirow}
\usepackage{algorithm}
\usepackage{algpseudocode}
\usepackage{pifont}
\usepackage{color}
\usepackage{subfigure}
\usepackage{enumitem}

\newcommand{\mylistbegin}{
  \begin{list}{$\bullet$}
   {
     \setlength{\itemsep}{-2pt}
     \setlength{\leftmargin}{1em}
     \setlength{\labelwidth}{1em}
     \setlength{\labelsep}{0.5em} } }
\newcommand{\mylistend}{
   \end{list}  }

\newcommand{\eg}{\textit{e.g.}}
\newcommand{\xeg}{\textit{E.g.}}
\newcommand{\ie}{\textit{i.e.}}
\newcommand{\etc}{\textit{etc}}

\newcommand{\wrt}{\textit{w.r.t.~}}

\setcopyright{rightsretained}
\acmDOI{10.475/123_4}

\acmISBN{123-4567-24-567/08/06}

\acmConference[]{}{}{} 
\acmYear{2018}
\copyrightyear{2018}

\acmPrice{15.00}

\begin{document}
\title{Graph Clustering with Dynamic Embedding}
%
%
%
\author{Carl Yang, Mengxiong Liu, Zongyi Wang, Liyuan Liu, Jiawei Han}
       \affiliation{
       \institution{University of Illinois, Urbana Champaign}
       \streetaddress{201 N. Goodwin Ave}
       \city{Urbana}
       \state{Illinois}
       \postcode{61801}
       }
       \email{jiyang3, mliu60, zwang195, ll2, hanj@illinois.edu}

\setlength{\dbltextfloatsep}{3pt plus 2pt minus 1pt}
\setlength{\dblfloatsep}{3pt plus 2pt minus 1pt}

\begin{abstract}
Graph clustering (or community detection) has long drawn enormous attention from the research on web mining and information networks. Recent literature on this topic has reached a consensus that node contents and link structures should be integrated for reliable graph clustering, especially in an unsupervised setting. However, existing methods based on shallow models often suffer from content noise and sparsity. In this work, we propose to utilize deep embedding for graph clustering, motivated by the well-recognized power of neural networks in learning intrinsic content representations. Upon that, we capture the dynamic nature of networks through the principle of influence propagation and calculate the dynamic network embedding. Network clusters are then detected based on the stable state of such an embedding. Unlike most existing embedding methods that are task-agnostic, we simultaneously solve for the underlying node representations and the optimal clustering assignments in an end-to-end manner. To provide more insight, we theoretically analyze our interpretation of network clusters and find its underlying connections with two widely applied approaches for network modeling. Extensive experimental results on six real-world datasets including both social networks and citation networks demonstrate the superiority of our proposed model over the state-of-the-art.
\end{abstract}

\maketitle
\section{Introduction}
\label{sec:intro}
One of the most popular topics in web and network research is to identify clusters. 
On the one hand, as the web is growing larger than ever before, it is necessary and efficient to look into smaller sub-networks, which consist of specific groups of interacting objects (\eg, web pages, users, papers) and their links (\eg, hyperlinks, friendships, citations). 
On the other hand, the knowledge of cluster structures allows us to better understand the status of an object within a group and the relations between it and its peers. This enables multiple benefits such as the discovery of functionally related objects \cite{streich2009multi}, the study of interactions between modules \cite{airoldi2008mixed}, the inference of missing attributes \cite{yang2017bi}, the prediction of unobserved connections \cite{wei2017cross} and so on. 

Thanks to the efforts of intensive recent research on network modeling \cite{liu2015community, wei2017cross, yang2017bi, yang2009combining, ruan2013efficient, chakrabarti2014joint, mcauley2012learning, yang2013community}, the importance of integrating contents and links in the networks for reliable community detection has been well recognized. 
In general, links in the networks are highly incomplete, \eg, friends in the real world do not always build links on a social platform, and semantically related publications do not always cite each other. In such scenarios, contents like user attributes and paper abstracts can help the retrieval of missing links. Upon this observation, many methods have been developed and shown to be effective to some extent, such as those based on probabilistic generative models \cite{zhou2012community, mcauley2012learning, yang2013community}, graph augmentation \cite{tang2015pte, yang2009combining, ruan2013efficient}, and information propagation \cite{liu2015community, chakrabarti2014joint, yang2017bi}.

However, as shown in Table \ref{tab:contents}, contents themselves in the networks are very sparse, \eg, users on social networks usually do not fill in many attributes, and papers only cover parts of all relevant words. Moreover, contents are also noisy, as users may fill in arbitrary fake information, and papers might be written with different terminology even for the same topics. Such sparse and noisy contents pose severe challenges to traditional models, as they become inefficient when estimating large volumes of parameters, and are ineffective in deeply understanding the complex intrinsic content distribution.

Recently, significant attention on network modeling has been paid to methods related to the progressive development of deep learning \cite{cho2014properties, mikolov2013distributed, krizhevsky2012imagenet, le2011learning}. Among most of them, the essential leverage is the computation of a network embedding, \ie, the distributed representations of objects that reflect the network structures \cite{perozzi2014deepwalk, tang2015line, grover2016node2vec, tang2015pte, yang2015network, cao2015grarep, wang2016structural, yang2016revisiting, niepert2016learning, li2015gated, kipf2016semi, defferrard2016convolutional, bruna2013spectral}. Some of them (\eg, \cite{perozzi2014deepwalk, tang2015line, grover2016node2vec, tang2015pte, yang2015network, cao2015grarep}) are based on shallow neural networks like Skip-gram \cite{mikolov2013distributed, levy2014neural}. They are effective in capturing network proximities and simple structures but do not work well with sparse noisy network contents. Other methods like \cite{wang2016structural, yang2016revisiting, niepert2016learning, li2015gated, kipf2016semi, defferrard2016convolutional, bruna2013spectral, wang2017topological} are built on more powerful deep neural networks like CNN \cite{krizhevsky2012imagenet, le2011learning} and RNN \cite{cho2014properties, wang2017topological}, for the better understanding of network contents and more complex structures, but supervision is usually required for them to be effective.

In this work, we propose GRACE (GRAph Clustering with dynamic Embedding), to address the challenging problem of unsupervised graph clustering through task-aware network embedding. 
The technique is motivated by the recent success in deep embedding for image and text processing \cite{xie2016unsupervised, guo2017improved}, and is then tailored for network clustering based on the principle of influence propagation \cite{yang2017bi, liu2015community}. 

Specifically, multiple nonlinear layers of deep denoise autoencoder with an embedding loss are devised to learn the intrinsic distributed representation of sparse noisy contents in networks. 
The deep content embedding is then propagated on the network through the principle of influence propagation, and network clusters are detected based on the stable state of such a dynamic network embedding. To further improve learning efficiency, we incorporate an unsupervised self-training clustering component to allow the framework to learn from its own high-confidence predictions and jointly train clustering and embedding in an end-to-end manner.

To provide more insight into network clusters and corroborate our model design, we theoretically analyze the principle of influence propagation. We demonstrate that our leverage of network dynamic embedding leads to a mathematical model that is essentially related to two major approaches for network modeling, but our model is more efficient without the need for sampling or supervision. 

We conduct extensive experiments on six real-world datasets including both social networks and citation networks. The performance of GRACE is supreme compared with various state-of-the-art network embedding and community detection algorithms.

The rest of this paper is organized as follows. Section 2 covers our model details and theoretical discussions. In Section 3, we present extensive experimental results including comparisons with various baselines as well as comprehensive studies of our GRACE framework. Related works are discussed in Section 4 and a quick summary is provided in Section 5.

\section{GRACE}
\label{sec:model}
The objective of this work is unsupervised clustering on networks with contents. We summarize the main challenges of this task into the following three aspects.
\begin{itemize}
\item Node contents are high-dimensional, sparse and noisy.
\item Network dynamics are crucial but hard to integrate.
\item No supervision is available to guide exploration.
\end{itemize}
To deal with all challenges above, we develop GRAph Clustering with dynamic Embedding (GRACE), which inherently combines deep embedding with influence propagation to integrate network node contents and link structures. A self-training clustering framework is adopted to further improve the performance.

\subsection{Overall Framework}
{\flushleft \textbf{Input.}}
We are given a network modeled by $\mathcal{G}=\{\mathcal{V},\mathcal{E},\mathcal{A}, \mathcal{C}\}$, where $\mathcal{V}$ is the set of $n$ nodes (\eg, users on social networks, papers in citation networks, \etc), $\mathcal{E}$ is the set of $m$ observed links among the nodes in $\mathcal{V}$. $\mathcal{A}$ is the set of observed node contents associated with $\mathcal{V}$, where each $\mathbf{a}_i$ is the set of $\kappa$ content features on node $v_i$. $\mathcal{C}$ is a set of ground-truth cluster memberships, where $\mathbf{c}_{ik}=1$ means node $v_i$ is in the $k$th ground-truth cluster. For unsupervised graph clustering, $\mathcal{C}$ is only used for evaluation.

{\flushleft \textbf{Output.}}
Our model jointly learns a deep embedding $\mathcal{X}$ and a soft clustering $\mathcal{Q}$ on $\mathcal{V}$. Each $\mathbf{x}_i \in \mathcal{X}$ is a distributed representation of node $v_i$, capturing a prominent nonlinear combination of its original features $\mathbf{a}_i$ under the consideration of $v_i$'s contexts in its neighborhood $\mathcal{N}_i$ on the network. The definition of $\mathcal{N}_i$ will be introduced in the following subsection (Section \ref{sec:model}.3), together with our specific treatment to $\mathcal{X}$ based on the principle of influence propagation on networks. Each $\mathbf{q}_i \in \mathcal{Q}$ is a $K$-dimensional probability distribution, where $K$ is the number of clusters to be detected and $\sum_k q_{ik}=1, \forall i$, characterizing the probability of node $v_i$ belonging to the $k$th cluster.

\begin{figure}[h!]
        \includegraphics[width=1\linewidth]{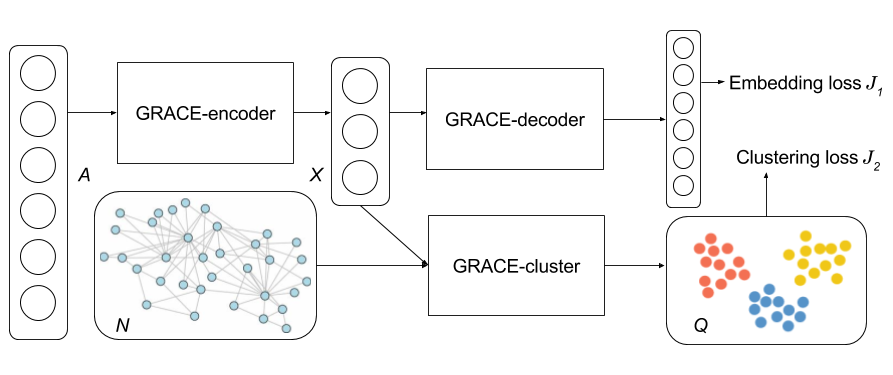}
    \caption{The overall architecture of GRACE.}
    \label{fig:overall}
 \end{figure}

{\flushleft \textbf{Architecture.}}
Figure \ref{fig:overall} illustrates the overall architecture of our GRACE framework.
We take the input $\mathcal{A}$ for all nodes $\mathcal{V}$ and compute their nonlinear deep embedding $\mathcal{X}$, which should be able to reconstruct the original node contents with an embedding loss $\mathcal{J}_1$. At the meantime, $\mathcal{X}$ is propagated among neighbors $\mathcal{N}$ on the network to generate the dynamic network embedding, which leads to robust graph clustering results $\mathcal{Q}$ with a clustering loss $\mathcal{J}_2$. Therefore, we have the overall loss function
\begin{align}
\mathcal{J}=\mathcal{J}_1+\lambda \mathcal{J}_2,
\label{eq:overall}
\end{align}
where $\lambda$ is a weighting parameter. By minimizing Eq.~\ref{eq:overall}, we can jointly optimize the deep embedding and soft clustering modules and allow them to mutually enhance each other in a closed loop.

In what follows, we further explain the reasons for our model architecture and how it works in details.

\subsection{Deep Embedding of Sparse Contents}
\begin{table*}[h]
\small
 \centering
 \begin{tabular}{|c|c|cccccc|}
 \hline
Type&Dataset&Nature of contents&\#Features&Avg. sparsity&\#Nodes&\#Links&\#Clusters\\
\hline
\multirow{2}{*}{\begin{tabular}{@{}c@{}} Social \\ Networks\end{tabular}}&Facebook \cite{mcauley2012learning}&0-1 vectors indicating the possession of attribute values&1,283&93.49\%&4,039&88,234&193\\
\cline{2-8}
&Gplus \cite{mcauley2012learning}&0-1 vectors indicating the possession of attribute values&15,907&98.85\%&107,614&13,673,453&468\\
\cline{2-8}
&Twitter \cite{mcauley2012learning}&0-1 vectors indicating the usage of hashtags&216,839&96.73\%&81,306&1,768,149&4,065\\
\hline
\multirow{2}{*}{\begin{tabular}{@{}c@{}}Citation \\ Networks\end{tabular}}&Cora \cite{sen2008collective}&0-1 vectors indicating the usage of words&1,433&98.73\%&2,708&5,429&7\\
\cline{2-8}
&Citeseer \cite{sen2008collective} &0-1 vectors indicating the usage of words&3,703&99.14\%&3,312&4,715&6\\
\cline{2-8}
&PubMed \cite{namata2012query} &real-valued vectors recording the TF-IDF of used words&500&89.98\%&19,717&44,338&3\\
\hline
 \end{tabular}
 \caption{\label{tab:contents}\textbf{Summary of statistics of 6 real-world network datasets. More details of them are introduced in Section \ref{sec:exp}.1.}}
\end{table*}
Table \ref{tab:contents} illustrates the nature, dimension and sparsity of contents in several real-world network datasets. Such high dimensionality and sparsity are not surprising, because users on social networks are known to be reluctant in filling out many attributes and papers in citation networks only cover their limited vocabularies. Moreover, users are free to fill in fake attributes and papers can contain different terminologies, which make contents in networks not only high-dimensional and sparse, but also noisy and inaccurate.

In our situation, it is desirable to capture attributes precisely, because that they often become the signatures of network clusters. \xeg, in a football fans' club, a popular player identified by the contents of his tweets is likely to be the center of a community, surrounded by fans posting their semantically related but different tweets.

However, we note that existing network embedding algorithms are ineffective with sparse noisy contents because they usually start from preserving the link structures among non-attributed nodes \cite{grover2016node2vec, tang2015line, perozzi2014deepwalk}, and then incorporate attributes as augmented nodes \cite{tang2015pte}, text feature matrices \cite{yang2015network} or bag-of-word vectors \cite{yang2016revisiting}. In such ways, attributes are shallowly modeled as auxiliary information and the deep semantics within them are not fully explored.

To deal with such a challenge, we get inspired by recent successes in deep learning for feature composition \cite{vincent2008extracting, le2013building}. Specifically, we are interested in leveraging deep embedding to discover the latent distributed representations of node contents in networks, which, in an ideal case, are low-dimensional, dense and robust to noise.

To this end, we employ deep denoise autoencoder (DAE) \cite{vincent2008extracting, le2013building}, which has been proven advantageous in capturing the intrinsic features within high-dimensional sparse noisy inputs in an unsupervised learning fashion. 

To leverage the power of DAE, given a node $v_i$'s original content vector $\mathbf{a}_i$, we firstly apply a GRACE-encoder, which consists of multiple layers of fully connected feedforward neural networks with Exponential Linear Unit (ELU) activations \cite{clevert2015fast}. The neural networks are in decreasing sizes and after them we get a $S$-dimensional latent representation $\mathbf{x}_i$ as
\begin{align}
\mathbf{x}_i=\mathbf{f}_e^H(\ldots \mathbf{f}_e^2(\mathbf{f}_e^1(\mathbf{a}_i))\ldots),
\end{align}
where
\begin{align}
\mathbf{f}_e^h(\mathbf{x}) = ELU(\mathbf{W}_e^h Dropout(\mathbf{x})+\mathbf{b}^h_e).
\end{align}
$H$ is the number of hidden layers in GRACE-encoder. $\Theta_e$ is the set of parameters in the $H$ encoder layers.

\begin{figure}[t]
        \includegraphics[width=1\linewidth]{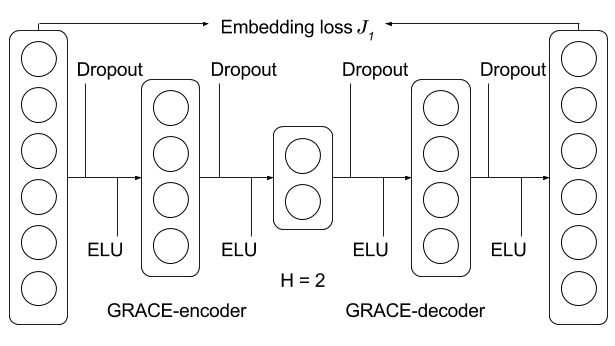}
    \caption{The deep embedding architecture of GRACE.}
    \label{fig:embed}
 \end{figure}
 
To ensure that $\mathbf{x}_i$ captures the important information in $\mathbf{a}_i$, we compute the reconstruction $\mathbf{\tilde{a}}_i$ of $\mathbf{a}_i$ through stacking a GRACE-decoder, which also consists of multiple layers of fully connected feedforward neural networks with ELU activations. The sizes of neural networks are in an increasing order, exactly the opposite as in GRACE-encoder.  So we have
\begin{align}
\mathbf{\tilde{a}}_i=\mathbf{f}_d^H(\ldots \mathbf{f}_d^2(\mathbf{f}_d^1(\mathbf{x}_i))\ldots),
\end{align}
where
\begin{align}
\mathbf{f}_d^h(\mathbf{x}) = ELU(\mathbf{W}_d^h Dropout(\mathbf{x})+\mathbf{b}^h_d).
\end{align}
The number of hidden layers in GRACE-decoder is also $H$, the same as in GRACE-encoder. $\Theta_d$ is the set of parameters in the $H$ decoder layers.

After GRACE-decoder, a reconstruction loss for embedding is computed as
\begin{align}
\mathcal{J}_1=\sum_{i=1}^n l(\mathbf{a}_i,\mathbf{\tilde{a}}_i),
\end{align}
which is a summation over all nodes in $\mathcal{V}$. Depending on the contents in the datasets, $l$ can be implemented either as a cross entropy (for binary features, such as user attributes) or a mean squared error (for continuous features, such as TF-IDF scores of words \cite{salton1986introduction}). 

Figure \ref{fig:embed} illustrates the deep embedding architecture of GRACE.
Its power mainly comes from the supreme expressiveness of deep feedforward neural networks \cite{hornik1989multilayer} with the randomness brought by the dropout technique \cite{srivastava2014dropout} that effectively alleviates overfitting, while enabling efficient exploration of important semantic patterns hidden in the sparse noisy contents.

\subsection{\large Graph Clustering with Dynamic Embedding}
The major difference between graph clustering and general clustering lies in the availability of additional structural information, such as users' friendships on social platforms and papers' citations in publication datasets. Such links, while providing crucial information not necessarily captured by node contents, are hardly leveraged by existing content embedding frameworks.

To reliably find clusters (or communities) on networks, we study the nature of clusters and model them under the consideration of network dynamics-- nodes on networks are constantly interacting with each other, sending and receiving influences among themselves. Inspired by \cite{liu2015community, chakrabarti2014joint, li2014user}, we assume that cluster is a consequence of such influence propagation. Accordingly, they should be modeled based on the consistency of propagated influences on the network when they reach a stable state. Unlike traditional influence propagation studies that model a single influence factor that corresponds to node activations \cite{goldenberg2001talk, granovetter1978threshold}, we assume $S$-dimensional latent factors propagated on the network, \ie, the dynamic network embedding, which jointly influence the original $\kappa$-dimensional node contents. Such an interpretation also naturally fits into our content embedding framework.

Considering the standard influence propagation model \cite{xiang2013pagerank}, we use an adjacency matrix $W$ to record the link structures in $\mathcal{G}$. For simplicity, we assume 0-1 weights and undirected links, while the model generalizes trivially to weighted directed graphs. To ensure that every node can receive influence from at least one node, we also assume every node can propagate to itself. So we have
\begin{align}
w_{ij} =
\begin{cases}
1, & \text{if } e_{ij} \in \mathcal{E} \text{, or } e_{ji} \in \mathcal{E} \text{, or } i=j,\\
0, & \text{otherwise. }
\end{cases}
\label{eq:w}
\end{align}
Given $W$, we define $D$ as a diagonal matrix with $d_{ii}=\sum_{j=1}^n w_{ij}, \forall i \in [1,n]$. So we have $T=D^{-1}W$ as the transition matrix on $\mathcal{G}$. Then $X^1 = TX$ is the dynamic embedding propagated through one step. With $\mathcal{N}_i^b$ defined as the neighborhood of $v_i$ with size $b$ ($\forall b>0$), which includes all nodes that are no more than $b$ steps from $v_i$ on the network, $\mathbf{x}^b_i$ corresponds to the embedding of $v_i$ under the consideration of $v_i$'s network contexts in $\mathcal{N}_i^b$.

As we assume, the formation of clusters requires the dynamic embedding to reach a stable state. It corresponds to infinite steps of propagation, or stationary as in random walk terminology. Therefore, we consider a stochastic description about the probability of a node influencing another with a linear approximation of influence propagation.

Following \cite{liu2015community}, we use $R$ to denote the stationary propagation matrix, where we denote the probability that the influence of node $v_i$ is propagated to node $v_j$and  as $r_{ij}$, which satisfies
\begin{align}
\forall i,j \in [1,n], i\neq j: \; r_{ii} = \beta + \alpha \sum_{e_{lj}\in \mathcal{E}}r_{il}t_{li},\; r_{ij} = \alpha \sum_{e_{lj}\in \mathcal{E}}r_{il}t_{lj},
\label{eq:r}
\end{align}
where $t_{ij}\in T$ is the transition probability from $v_i$ to $v_j$, $\beta>0$ is a constant corresponding to the probability of a node propagating its contents to itself, $\alpha$ is the damping coefficient of the propagation process.

Let $\gamma=\frac{\alpha}{\beta}$. According to the constraints in Eq.~\ref{eq:r}, we can compute $R$ as
\begin{align}
R=(\beta I-\alpha T)^{-1}=\beta(I-\gamma D^{-1}W)^{-1}.
\label{eq:rmat}
\end{align}
Subsequently, we have
\begin{align}
\tilde{X}=X^{inf}=RX=\beta(I-\gamma D^{-1}W)^{-1}X,
\end{align}
where $\beta$ can be simply removed because it is the same for every node and does not affect the clustering results.

Describing network clusters with stable influence propagation, we integrate the leverage of link structures into GRACE by substituting the deep embedding $X$ with the dynamic embedding $\tilde{X}$ before clustering computation. We defer the design of clustering loss $\mathcal{J}_2$ to the following subsection (Section \ref{sec:model}.4).

\begin{figure}[h!]
        \includegraphics[width=1\linewidth]{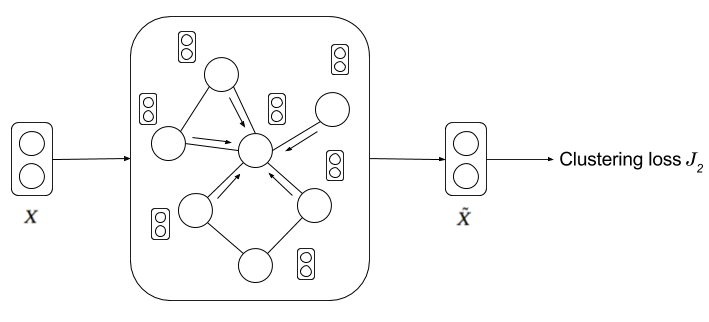}
    \caption{\small The dynamic embedding-based clustering of GRACE.}
    \label{fig:cluster}
 \end{figure}

Figure \ref{fig:cluster} illustrates our clustering component with network dynamics. A clustering loss $\mathcal{J}_2$ is computed based on $\tilde{X}$ and back-propagated to the embedding model via $R$, which guides the deep exploration of contents in $\mathcal{A}$ on the network. By doing this, we require the embedding of two nodes to be close not only because they share similar important contents in a compact latent space, but also because they send/receive similar influence to/from their neighboring nodes on the network.


{\flushleft \textbf{Connections to other popular methods.}}
To better understand the nature of network clusters and corroborate the insight and design of dynamic embedding through influence propagation, we theoretically explore its connections to popular related algorithms for network modeling.
To this end, we note that there exist two major well-known methods for integrating network dynamics into embedding models, \ie, graph Laplacian based regularization \cite{yang2016revisiting, yang2017bridging} and spectral graph convolution \cite{kipf2016semi, defferrard2016convolutional}. 

The first approach leverages the idea of context preservation introduced by \cite{perozzi2014deepwalk} into a semi-supervised learning framework for regularizing node embedding \wrt network structures \cite{yang2016revisiting, yang2017bridging}. It is shown in \cite{yang2017bridging} that such a context preservation objective based on random walk sampling on networks is a generalization of traditional graph Laplacian regularizer, which basically introduces a regularization term in the following form
\begin{align}
\mathcal{J}_r=\sum_{i,j\in [1,n]} w_{ij} || \mathbf{x}_i-\mathbf{x}_j||^2 = X^T L^{sym} X,
\end{align}
where $w_{ij} \in W$ is the same as in Eq.~\ref{eq:w}, and $L^{sym}$ is the symmetric normalized graph Laplacian matrix, \ie, $L^{sym}=I-D^{-1/2}WD^{-1/2}$. Their underlying intuition matches our influence propagation model-- to require nodes close on the network to have similar latent representations, as we discuss in Section \ref{sec:model}.3. However, to approximate $\mathcal{J}_r$, their approach requires additional computation of random walk sampling and the sampling process itself usually needs significant tuning \cite{chen2017sampling}.

The second approach captures network dynamics through convolution on graphs with localized spectral filters \cite{kipf2016semi, defferrard2016convolutional}. They essentially leverage the graph Laplacian matrix to compute the graph Fourier transform of node contents as
\begin{align}
\tilde{X}=g_\theta(L^{sym})X=g_\theta(U\Lambda U^T)X=Ug_\theta(\Lambda)U^TX,
\label{eq:spectral}
\end{align} 
where $g_\theta$ is a function of the eigenvalues $\Lambda$ of $L^{sym}$, and $U$ is the matrix of the eigenvectors of $L^{sym}$. 

To avoid the eigen-decomposition of $L^{sym}$, $\tilde{X}$ is usually approximated with a truncated expansion in terms of Chebyshev polynomials up to $T$th order \cite{hammond2011wavelets}. As derived in \cite{kipf2016semi}, with $T$ set to 1 and some other constraints on the Fourier coefficients, Eq.~\ref{eq:spectral} can be simplified into
\begin{align}
\tilde{X}\approx\phi(I+D^{-1/2}WD^{-1/2})X,
\end{align}
where $\phi$ are the learnable Fourier coefficients.

Note that, assuming $W$ is without self-loop, this approximated Fourier transformation matrix $(I+D^{-1/2}WD^{-1/2})$ is very close to our one-step transition matrix $T$ defined based on Eq.~\ref{eq:w}. The difference is that we add self-loops before normalizing the adjacency matrix, and they add self-loops after normalizing the same matrix. 
Therefore, according to \cite{hammond2011wavelets}, our modeling of influence propagation is similar to applying multiple 1-localized spectral filters. It corresponds to linear graph convolutions on the graph Laplacian spectrum and is an effective way of capturing complex network dynamics. The difference is that our influence propagation model is purely unsupervised and much more efficient without the need for learning the Chebyshev coefficients $\theta$.

\subsection{Jointly Learning Clustering and Embedding}
The final challenge of clustering on graphs lies in the lack of labeled data, \ie, the unsupervised learning scenario. Most successful deep learning frameworks on image or language processing are usually composed of an unsupervised pre-training step and a supervised fine-tuning step so that the intrinsic structures of data and supervision can be combined to automatically capture meaningful visual or semantic patterns. However, graph clustering is naturally unsupervised, which means although our deep neural networks can be powerful in exploring various patterns, they can not receive feedback from the environment and can never explicitly know what patterns they find are useful.

In order to address this challenge, we get inspired by recent progress on self-training neural networks. The idea is to let the model iteratively make predictions and learn from its own high confidence predictions, which in turn helps improve the low confidence predictions \cite{nigam2000analyzing}.
Accordingly, we jointly train clustering and embedding by borrowing the idea of minimizing the KL divergence between a soft clustering distribution $\mathcal{Q}$ and an auxiliary target distribution $\mathcal{P}$ \cite{xie2016unsupervised, guo2017improved}, which was originally designed to cluster images and documents without the network setting.

Specifically, after getting the node embedding $\mathcal{X}$, besides reconstructing node contents through GRACE-decoder, we input $\mathcal{X}$ (or $\mathcal{\tilde{X}}$ after influence propagation) into a GRACE-cluster module, which computes soft clustering assignments $\mathcal{Q}$ to all nodes in $\mathcal{V}$ according to the following equation
\begin{align}
q_{ik} = \frac{(1+||\mathbf{x}_i-\mathbf{u}_k||^2)^{-1}}{\sum_j(1+||\mathbf{x}_i-\mathbf{u}_j||^2)^{-1}}.
\end{align}
In this equation, $q_{ik}$ is the probability of assigning node $v_i$ to the $k$th cluster, under the assumption of Student's $t$-distribution with degree of freedom set to 1 \cite{maaten2008visualizing}. $q_{ik}$ is basically a kernel function that measures the similarity between the embedding of node $v_i$ and the cluster center $u_k$.

The idea of self-training is to iteratively learn from more confident clustering results. Following \cite{xie2016unsupervised, guo2017improved}, we use an auxiliary target distribution $\mathcal{P}$ defined as
\begin{align}
p_{ik} = \frac{q_{ik}^2/f_k}{\sum_j q_{ij}^2/f_j},
\label{eq:p}
\end{align}
where $f_k=\sum_i q_{ik}$ is the total number of nodes softly assigned to the $k$th cluster. Raising $q$ to the second power and then dividing by the frequency per cluster allows the target distribution $\mathcal{P}$ to improve cluster purity and stress on confident assignments, while normalizing the contribution of each centroid on the clustering loss to prevent large clusters from distorting the hidden feature space.

Our clustering loss is then defined as the KL divergence between the soft assignment distribution $\mathcal{Q}$ and the auxiliary target distribution $\mathbf{P}$ as following
\begin{align}
\mathcal{J}_2=KL(\mathcal{P}||\mathcal{Q})=\sum_i\sum_k p_{ik} log \frac{p_{ik}}{q_{ik}}.
\end{align}

\begin{figure}[h!]
        \includegraphics[width=0.9\linewidth]{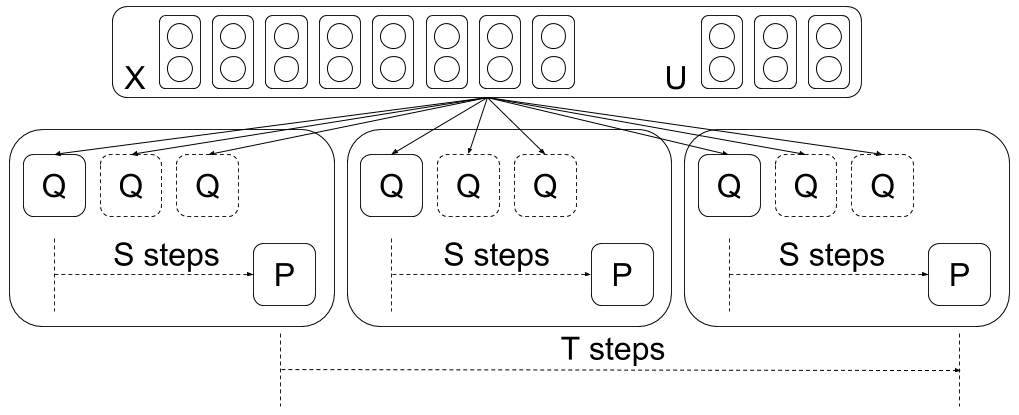}
    \caption{The joint self-training process of GRACE.}
    \label{fig:joint}
 \end{figure}

Figure \ref{fig:joint} illustrates this self-training process of GRACE. During the process, at each macro-step $t\;(0<t<T)$, we compute $\mathcal{P}$ through Eq.~\ref{eq:p} and fix it. Then we jointly update the node embedding $\mathcal{X}$ and the cluster centers $\mathcal{U}$ for $S$ micro-steps, which as a consequence updates the clustering distribution $\mathcal{Q}$. Therefore, the whole self-training process is like a chase-and-run game. During each of the $T$ total steps, $\mathcal{P}$ takes a large step based on the current value of $\mathcal{Q}$, and then $\mathcal{Q}$ takes $S$ small steps to catch up with $\mathcal{P}$, making improvements on both the clustering $\mathcal{Q}$ itself and the embedding $\mathcal{X}$ along the way.

\subsection{Training, Implementation, and Scalability}
We achieve the joint learning of clustering and embedding by combining pre-training and co-training.
Initialization is known to be crucial for most clustering algorithms. Therefore, before assigning nodes to clusters for the first time, we pre-train our GRACE-encoder and GRACE-decoder without GRACE-cluster for $T_0$ iterations or until the embedding loss $\mathcal{J}_1$ is sufficiently low. Then we run $K$-means on the pre-trained embedding $\mathcal{X}^0$ to initialize the cluster centers $\mathcal{U}^0$. Next, we plug in GRACE-cluster, and simultaneously optimize the embedding loss $\mathcal{J}_1$ and clustering loss $\mathcal{J}_2$ through co-training for $T$ iterations or until both losses are sufficiently low. As discussed in Section \ref{sec:model}.4, within each of the $T$ macro-steps, we compute $\mathcal{P}$ at the beginning and then use $S$ micro-steps during which we update other parts of the model with $\mathcal{P}$ fixed. The value of $\mathcal{Q}$ after the last update records the final cluster assignments and can be used for clustering prediction and evaluation.

To optimize Eq.~\ref{eq:overall}, we employ stochastic gradient descent (SGD) with mini-batch Adam. At the every step, the parameters $\Theta$ ($\Theta=\{\Theta_e, \Theta_d\}$ for pre-training and $\Theta=\{\Theta_e, \Theta_d, \mathcal{Q}\}$ for co-training) is updated by
\begin{eqnarray}
\Theta_{t} \gets \Theta_{t-1}-\frac{\rho}{\sqrt{\sum_{i=1}^{t}g_{i}^{2}}}g_{t},
\end{eqnarray}
where $\rho$ is the initial learning rate and $g_{t}$ is the sub-gradient at time $t$. The influence propagation model is simply implemented as a matrix multiplication that requires no training. The gradients are efficiently back-propagated to the embedding layers, and the overall learning framework is end-to-end.

We implement GRACE using TensorFlow\footnote{https://www.tensorflow.org/}, which runs efficiently on GPU. The codes will be available upon acceptance of the work.

Besides training the neural networks, we need to compute the inversion of a large $n\times n$ matrix for influence propagation, which is quite time consuming. Fortunately, this computation needs to be done only once before training and can be further approximated by multiplying $T$ for multiple times. 
In Section \ref{sec:exp}.3, we will empirically verify the approximation power of this approach.

\section {Experiments}
\label{sec:exp}
In this section, we evaluate GRACE for graph clustering with extensive experiments on 6 real-world network datasets.

\subsection{Experimental Settings}
{\flushleft \bf Datasets.}
We use six real-world network datasets that contain both node contents and link structures. In social networks, nodes correspond to users and links correspond to observed user connections such as friendships on Facebook and followings on Twitter. Each social network datasets include multiple ego-networks, so all compared algorithms are run on each of them with the final performance averaged over all separate runs. In citation networks, nodes are scientific publications (papers) and links are generated based paper citations. 
A summary of statistics of these datasets are presented in Table \ref{tab:contents} in Section \ref{sec:model}.2. 

{\flushleft \bf Compared algorithms.} 
We compare with two groups of algorithms from the state-of-the-art to comprehensively evaluate the performance of GRACE.
{\flushleft \it Network Community detection algorithms.} Some classic algorithms are based on network links alone, while more recent ones also leverage node contents. We compare with them to show the power of GRACE from deep embedding and influence propagation.
\begin{itemize}
\item \textbf{MinCut} \cite{clauset2004finding}: a classic community detection algorithm based on modularity.
\item \textbf{CESNA} \cite{yang2013community}: a generative model of edges and attributes to detect communities.
\item \textbf{PCL-DC} \cite{yang2009combining}: a unification of a conditional model for link analysis and a discriminative model for content analysis.
\item \textbf{CP} \cite{liu2015community}: a series of state-of-the-art community detection methods based on content propagation. We compare with the best variant called CPIP-SI.
\end{itemize}
{\flushleft \it Network node embedding algorithms.} While we find unsupervised node embedding helpful in capturing both link structures and node contents, we compare with the state-of-the-art embedding algorithms to show that GRACE is advantageous for clustering networks, especially those with sparse noisy contents.
The embedding learned by these algorithms are fed into the same $k$-means clustering algorithm to produce network clustering results.
\begin{itemize}
\item \textbf{DeepWalk} \cite{perozzi2014deepwalk}: an embedding algorithm based on truncated random walks that only uses network structures.
\item \textbf{TADW} \cite{yang2015network}: an embedding algorithm that generalizes DeepWalk to consider both node attributes and network structures by tensor factorization.
\item \textbf{PTE} \cite{tang2015pte}: an embedding algorithm that generalizes another powerful  algorithm called LINE \cite{tang2015line} to consider networks with contents by graph augmentation.
\item \textbf{NRCL} \cite{wei2017cross}: an embedding algorithm that aims to bridge the information gap by learning a robust consensus for link-based and attribute-based network representations.
\end{itemize}
The number of clusters to detect is tuned via standard 5-fold cross validation for all algorithms. 
The implementations of all compared algorithms are provided by their original authors, except for MinCut, which is provided in Stanford SNAP project\footnote{http://snap.stanford.edu/snap/index.html}.
{\flushleft \bf Metrics.}
Two widely used metrics for evaluating clustering results are used in our experiments.
For a detected cluster $c_i^*$ and a ground-truth cluster $c_i$, the \textit{F1 similarity} ($F1$) and \textit{Jaccard similarity} ($JC$) are defined as
\begin{align}
F1(c_i, c^*_i) = \frac{2 \cdot prec(c_i, c^*_i) \cdot rec(c_i, c^*_i)}{prec(c_i, c^*_i) + rec(c_i, c^*_i)},\; JC(c_i, c^*_i) = \frac{|c_i \cap c^*_i|}{|c_i \cup c^*_i|},
\end{align}
where $prec(c_i, c^*_i) = \frac{c_i \cap c^*_i}{|c^*_i|}$, $rec(c_i, c^*_i) = \frac{c_i \cap c^*_i}{|c_i|}$. 

For a set of detected clusters $C^*=\{c^*_i\}_{i=1}^M$ and a set of ground-truth clusters $C=\{c_i\}_{i=1}^N$, we compute the scores as
\begin{align}
F1(C, C^*) &= \sum_{c^*_i \in C^*} \frac{|c^*_i|}{\sum_{c^*_i\in C^*}|c^*_i|} \max_{c_i \in C} F1(c_i, c^*_i),\nonumber
\end{align}
\begin{align}
JC(C, C^*) & = \sum_{c_i \in C}\frac{\max_{c^*_i\in C^*} JC(c_i, c_i^*)}{2|C|}+\sum_{c^*_i \in C^*}\frac{\max_{c_i\in C} JC(c_i, c_i^*)}{2|C^*|}
\end{align} 

\subsection{Performance Comparison with Baselines}
We quantitatively evaluate GRACE against all baselines on network clustering.
To observe significant differences in performance, we run each non-deterministic algorithm 10 times to record the means and standard deviations. For deterministic algorithms, we run 1 time and treat the standard deviations as 0. Then we conduct paired statistical t-tests by putting GRACE against all baselines.

\begin{table*}[h]
 \centering
 \begin{tabular}{|c|cc|cc|cc|cc|cc|cc|}
 \hline
\multirow{2}{*}{Algorithm}&\multicolumn{2}{c|}{Facebook}&\multicolumn{2}{c|}{Gplus}&\multicolumn{2}{c|}{Twitter}&\multicolumn{2}{c|}{Cora}&\multicolumn{2}{c|}{Citeseer}&\multicolumn{2}{c|}{PubMed}\\
\cline{2-13}
&F1&JC&F1&JC&F1&JC&F1&JC&F1&JC&F1&JC\\
\hline
MinCut&0.2272&0.0731&0.2212&0.1355&0.2956&0.0970&0.4876&0.2897&0.4189&0.2144&0.5786&0.3419\\
\hline
CESNA&0.4532&0.3019&0.2057&0.1989&0.2976&0.1970&0.4244&0.2115&0.3643&0.1321&0.4612&0.2724\\
\hline
PCL-DC&0.4621&0.2554&0.2373&0.2156&0.3058&0.2062&0.5459&0.3723&0.5041&0.3227&0.5848&0.3978\\
\hline
CP&0.6248&0.3605&0.1450&0.1027&0.2083&0.1580&0.6921&0.5192&0.6912&0.4959&0.7017&0.5212\\
\hline
\hline
DeepWalk&0.4747&0.1432&0.2014&0.2027&0.2845&0.1859&0.6172&0.4402&0.4320&0.2649&0.6440&0.4808\\
\hline
TADW&0.5106&0.2841&0.2132&0.1886&0.2476&0.1526&0.5915&0.4142&0.4682&0.2807&0.6483&0.4897\\
\hline
PTE&0.2135&0.1184&0.1697&0.0671&0.1880&0.1016&0.3848&0.2526&0.2807&0.1366&0.4876&0.2410\\
\hline
NRCL&0.4938&0.2135&0.2254&0.2097&0.2988&0.2015&0.5219&0.3283&0.4439&0.2794&0.6280&0.4898\\
\hline
\hline
\textbf{GRACE}&\textbf{0.7212}&\textbf{0.4436}&\textbf{0.3110}&\textbf{0.2559}&\textbf{0.4132}&\textbf{0.2620}&\textbf{0.7428}&\textbf{0.5579}&\textbf{0.7397}&\textbf{0.5362}&\textbf{0.7378}&\textbf{0.5766}\\
\hline
 \end{tabular}
 \caption{\label{tab:perform}\textbf{Performance comparison with two groups of baselines on six real network datasets.}}
 \vspace{-10pt}
\end{table*}

The parameters of baselines are all tuned to the best through cross-validation. For GRACE, we empirically set the weight of clustering loss $\lambda$ to $0.1$; for the embedding model, we set the number of hidden layers $H$ to 2, with the first layer being half of the size as the original contents ($\kappa$) and following layers halving the sizes (\ie, $\kappa\to \kappa/2\to \kappa/4$ and $\kappa/4\to \kappa/2\to \kappa$ for the encoder and decoder, respectively); dropout rate is set to $0.5$; for the influence propagation model, we set the damping factor $\alpha$ to $0.9$, stressing on link structures in small local neighborhoods; for joint training embedding and clustering, we set the number of epochs $T_0=1000$, $T=30$ and $S=30$ for pre-training and co-training, respectively. As we will show in the following subsection, GRACE is not very sensitive to the setting of parameters and no much tuning is needed to achieve satisfactory performance.

Table \ref{tab:perform} shows the mean $F1$ and $JC$ scores evaluated for all compared algorithms. The results all passed the significant t-tests with $p$-value 0.01.
GRACE constantly outperforms all 8 compared algorithms by large margins on all 6 datasets in both F1 and Jaccard scores, while baselines have varying performances. This indicates the robustness and general advantages of GRACE. 

Taking a closer look, we observe that the advantage of GRACE over baselines is more significant on the social network datasets, where node contents are noisier. Among the citation networks, GRACE outperforms baselines more on Cora and Citeseer, the node contents of which are much sparser than PubMed. These two facts indicate the effectiveness of the GRACE deep embedding model in dealing with noisy sparse contents. 

For both community detection and network embedding baselines, combining node contents with link structures usually lead to better performance. However, this is not always true (\eg, consider MinCut \textit{vs.}~CESNA on citation networks, DeepWalk \textit{vs.}~TADW on Twitter and Cora). It implies the importance and correctness of our interpretation of clusters as a consequence of network dynamics and subsequently confirms our appropriate integration of contents and links via the influence propagation model.

All experiments are done on a server with four GeForce GTX 1080 GPUs and a 12-core 2.2GHz CPU. The runtime of GRACE is comparable to most baselines on CPU, while it runs $10+$ times faster on GPU.

\subsection{Deep Analysis of GRACE}
We comprehensively analyze the performance of GRACE with different content embedding and influence propagation architectures. The results confirm our insights about sparse noisy contents and network dynamics, while also justify our design of deep embedding and stationary propagation models. We also visualize the changes in the embedding spaces as we train the models to show how GRACE is able to jointly leverage node contents and link structures for capturing network clusters.

\subsubsection{Embedding Size and Depth} 
As we believe deep embedding is crucial for dealing with sparse noisy contents, we explicitly study the impact of embedding architectures by varying the size $S$ and depth $H$ of GRACE-encoder and GRACE-decoder. Specifically, we vary the embedding size $S$ from 256 for datasets with lower dimensions of original contents (Facebook and PubMed) and $512$ for the rest, to a small number like $16$ or $32$, to understand the content complexity of each dataset and how compact the useful representations can be. Each time, we halve the embedding dimension by 2. 
For each dataset, we vary the embedding depth $H$ from 1 to 4, to see how deep neural networks can help to effectively capture the compact representations. For simplicity, we do not use hidden layers of decreasing or increasing sizes as in Section \ref{sec:exp}.2. Instead, we set the sizes of all hidden layers to be the same. \xeg, with $S=256$ and $H=2$, the encoder and decoder have the architectures of $\kappa\to 256\to 256$ and $256\to 256\to \kappa$, respectively. 

\begin{figure*}[h!]
\centering
\subfigure[Facebook]{
\includegraphics[width=0.28\textwidth]{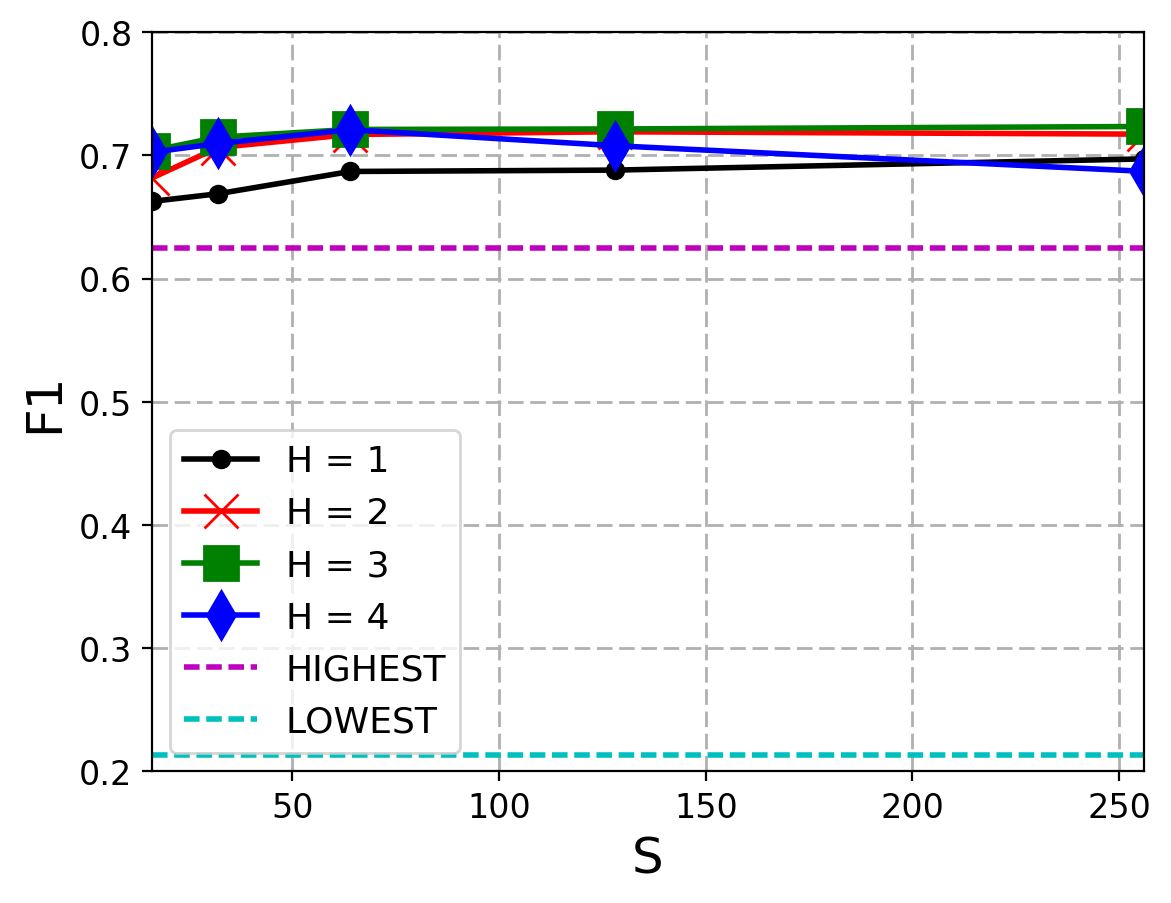}}
\subfigure[Gplus]{
\includegraphics[width=0.28\textwidth]{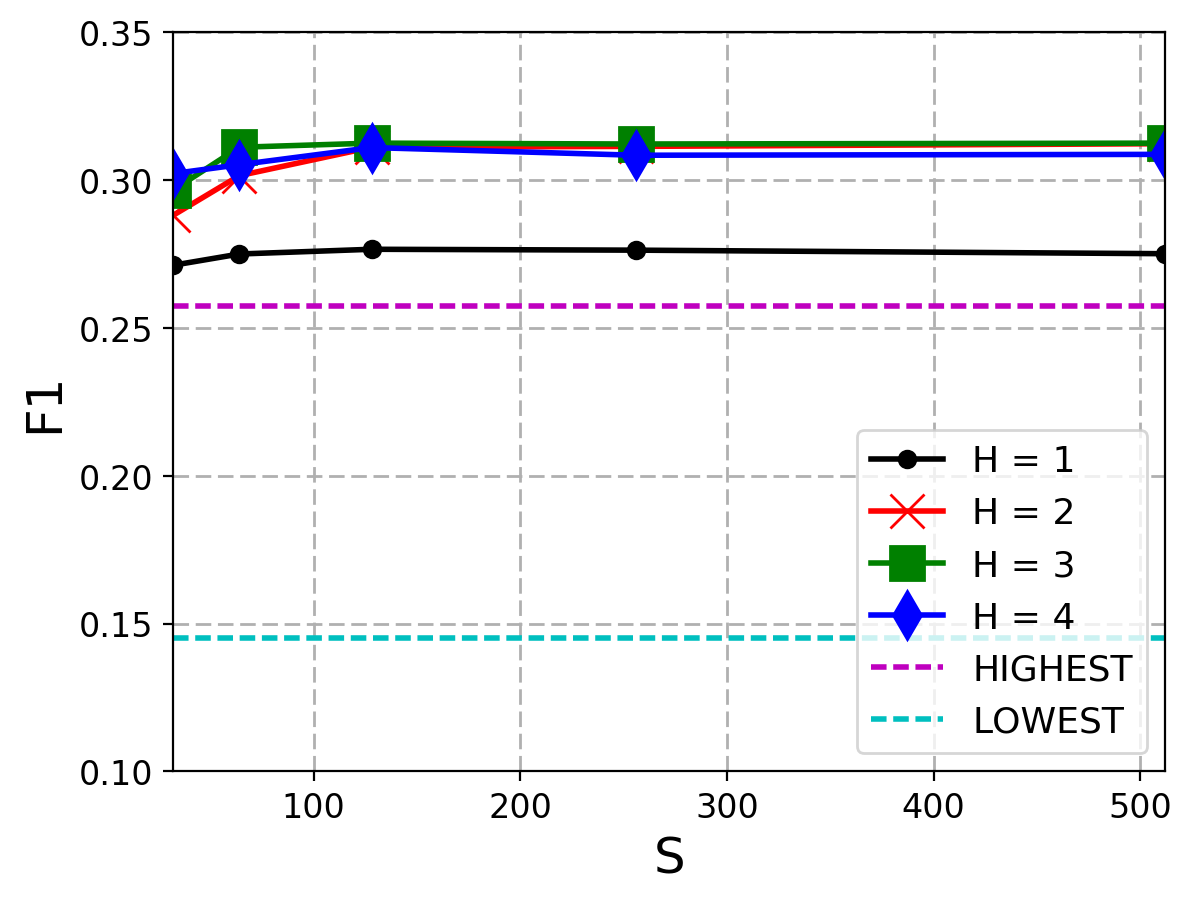}}
\subfigure[Twitter]{
\includegraphics[width=0.28\textwidth]{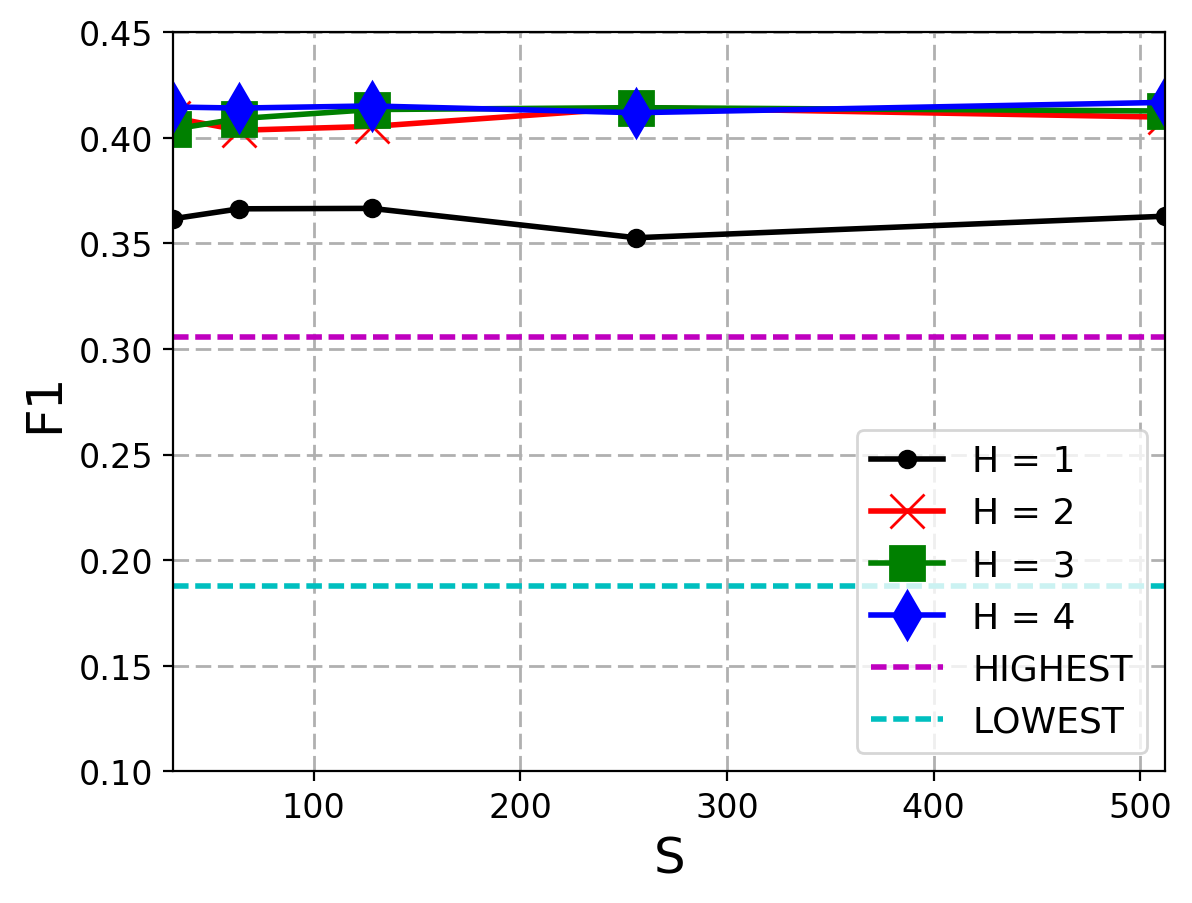}}
\subfigure[Cora]{
\includegraphics[width=0.28\textwidth]{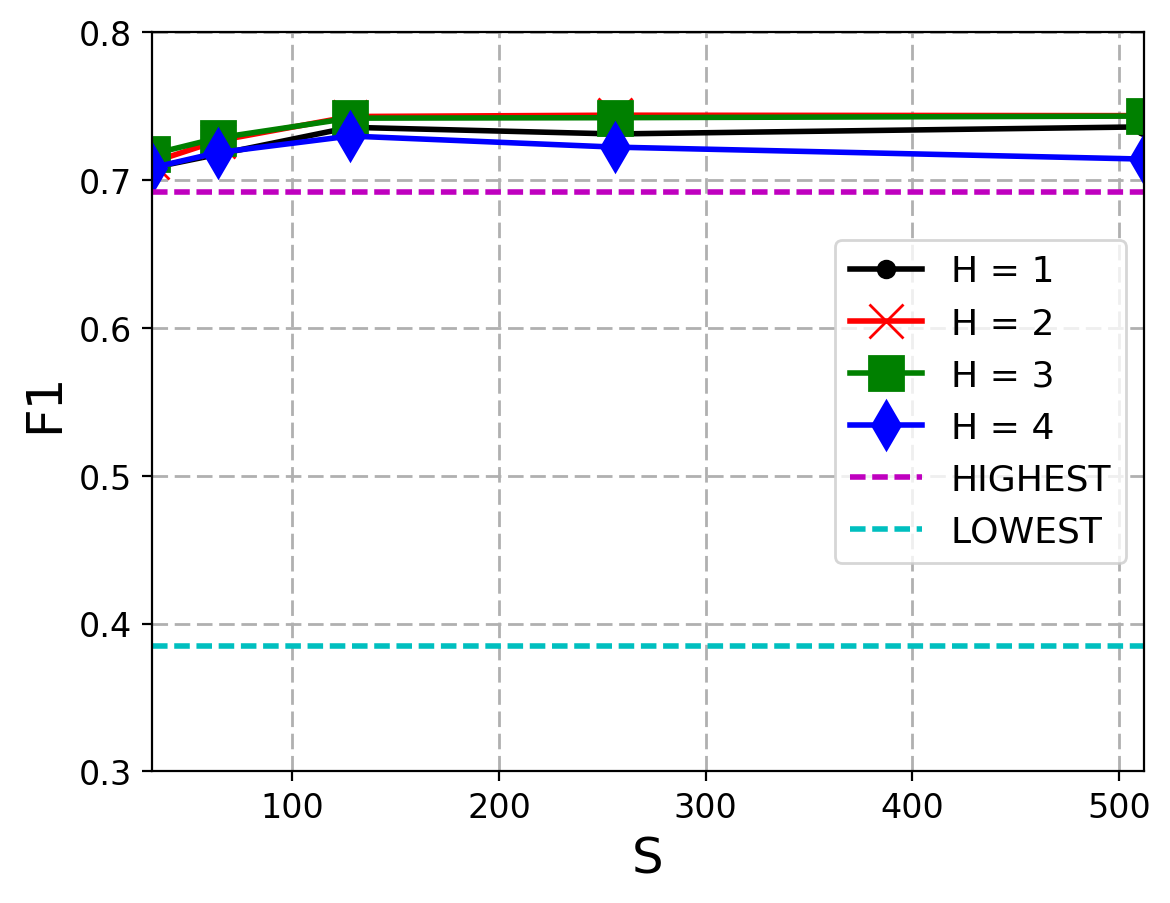}}
\subfigure[Citeseer]{
\includegraphics[width=0.28\textwidth]{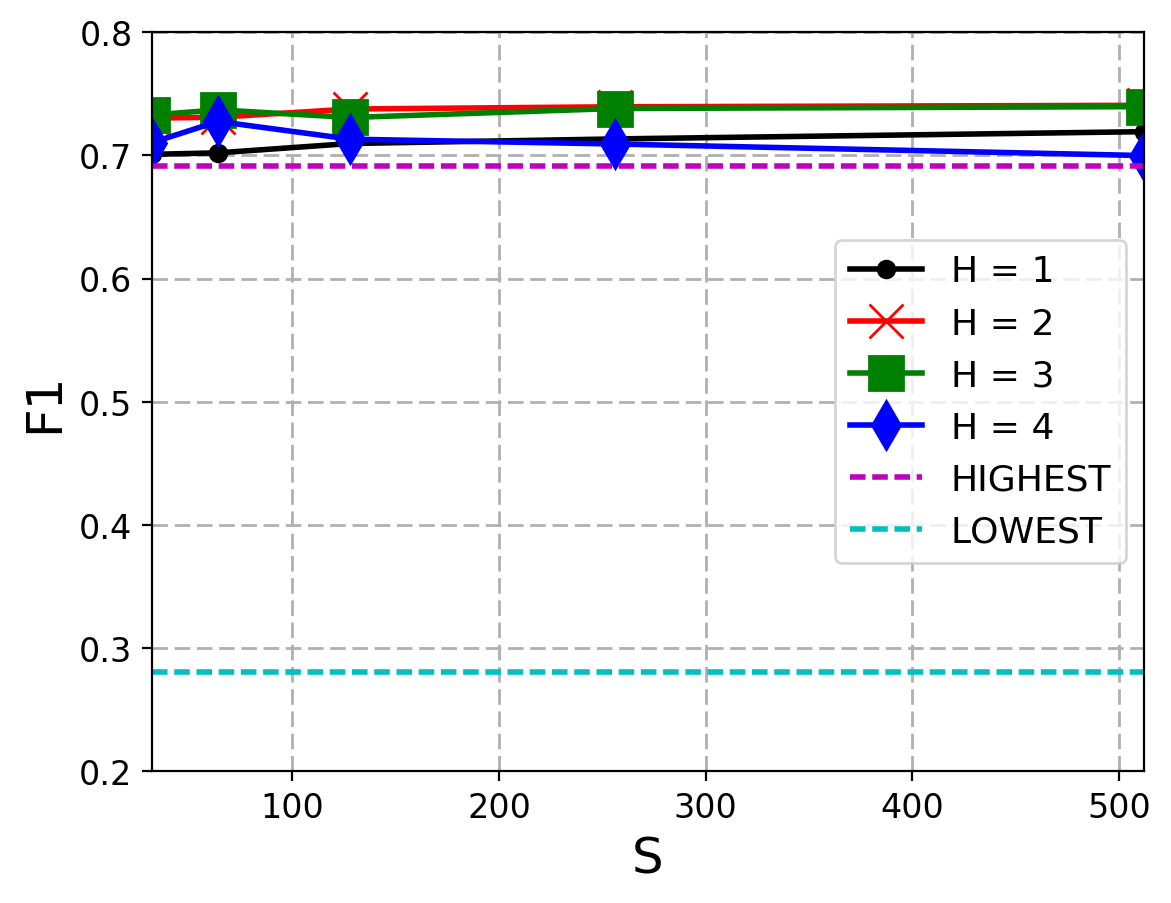}}
\subfigure[PubMed]{
\includegraphics[width=0.28\textwidth]{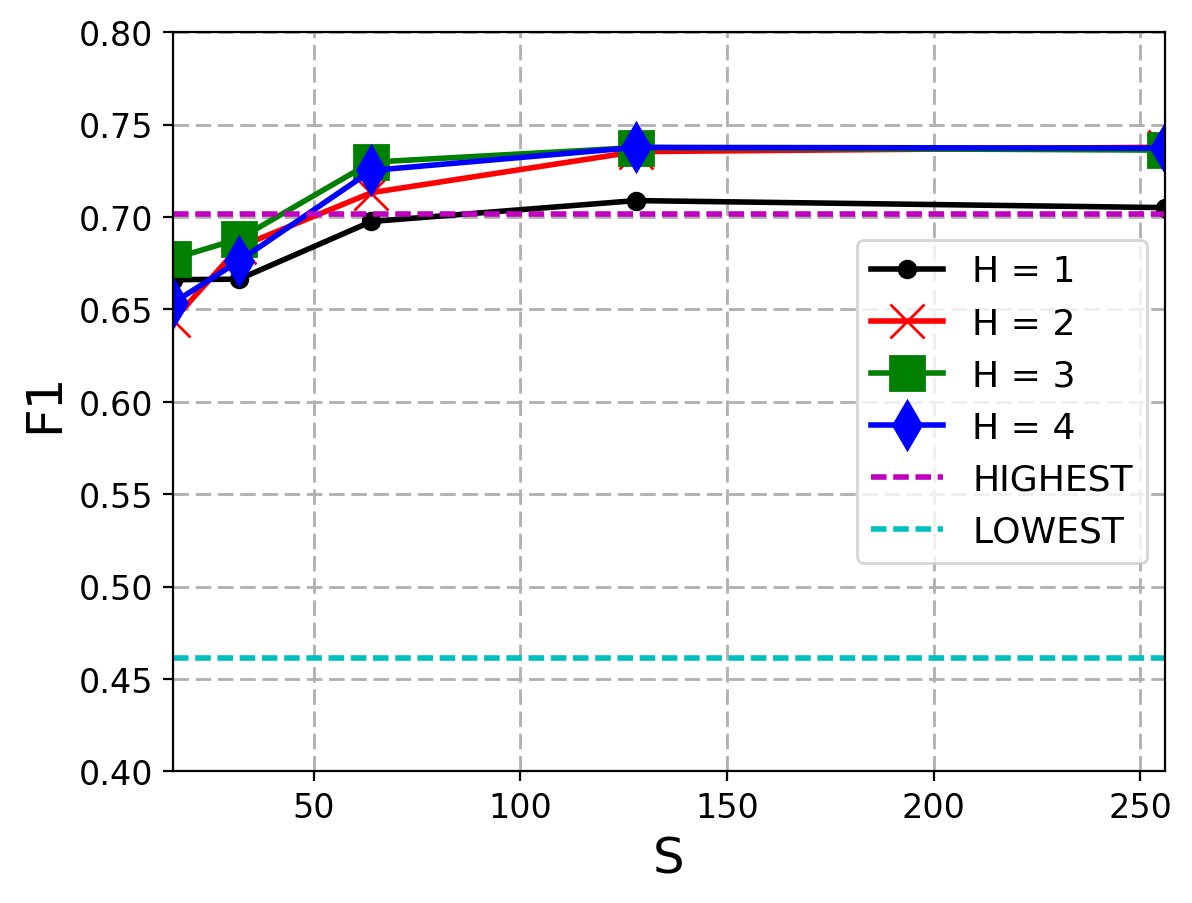}}
\vspace{-10pt}
\caption{F1 scores of model selection experiments on embedding size and depth.}
\label{fig:exp-emb}
\end{figure*}

Figure \ref{fig:exp-emb} shows the $F1$ scores on 6 datasets. The $JC$ scores show similar trends and are omitted due to the space limitation. For Facebook and PubMed, the ranges of $x$-ticks are smaller, because their original contents are of lower dimensions. 
As we can observe, the embedding size $S$ does not have a large impact on the performance of GRACE, except for the cases where it is too small for the embedding to capture the complex content semantics. This happens apparently on PubMed (probably because its contents are less sparse and more complex), and slightly on Facebook, Gplus and Cora, indicating their contents to be also more complex than those of Twitter and Citeseer. Setting $S$ to a value like half of the size of the original contents will usually lead to satisfactory results.

As for the embedding depth $H$, shallower neural networks with $H=1$ perform significantly worse than their deeper competitors on most datasets, \ie, Facebook, Gplus, Twitter and PubMed. Three out of the four datasets are social network datasets that are known to have noisier contents, and the last one PubMed has more complex contents as we have just discussed. This clearly confirms our insight of leveraging the power of deep embedding to effectively capture network node contents. Since we apply dropout in both encoder and decoder, only slight overfitting is observed on a few datasets that have relatively small networks (Facebook, Cora, and Citeseer), when the embedding depth is too large ($H=4$). Setting $H$ to a value like 2 or 3 will likely produce stable results.

\subsubsection{Stable Propagation and Its Approximations}
In this work, we understand network clusters as a consequence of influence propagation on networks, and we believe the cluster structures are most significant when such propagation among nodes reaches stability, \ie, the stationary distribution. In Eq.~\ref{eq:rmat}, we compute the stationary transition matrix $R$, which involves the inversion of a large matrix. Here, we simply use $T^b$, the multiplications of $T$, as an approximation of $R$, to show how well we can capture network clusters without reaching the stationary distribution.
 
\begin{figure*}[h!]
\centering
\includegraphics[width=1\textwidth]{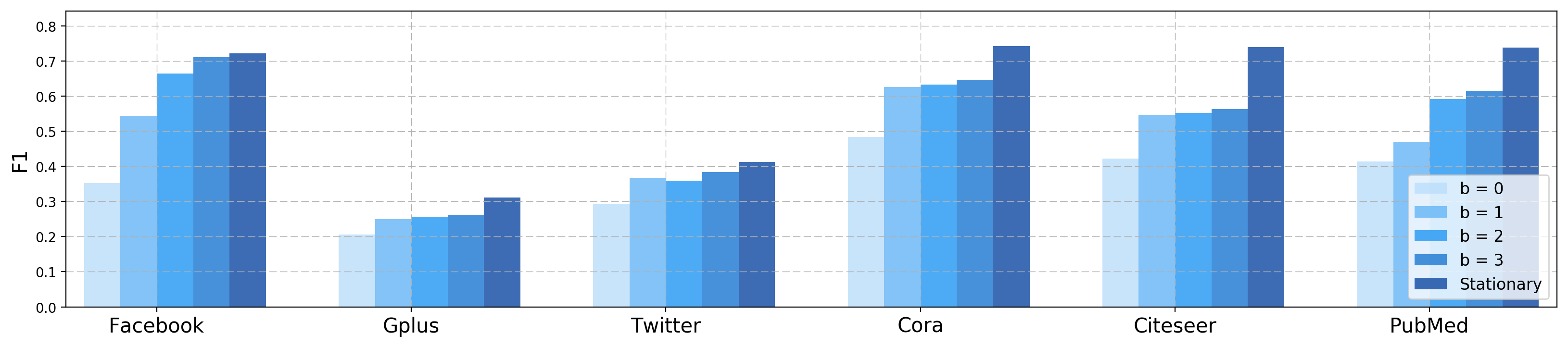}
\vspace{-20pt}
\caption{F1 scores of model selection experiments on stable influence propagation and its approximations.}
\label{fig:exp-prop}
\end{figure*}

As we can observe from Figure \ref{fig:exp-prop},
the number of propagation steps has a large impact on the performance of GRACE, especially when the number is small. This demonstrates the utility of our influence propagation model. As the number of steps grows large and an approximated stationary distribution is reached, the performance becomes stable. However, there is still a performance gap between the approximated and the true stationary distributions, because the damping effect cannot be trivially approached through multiplying one-step transition matrices, but cluster structures are often observed within small network neighborhoods. 

\subsubsection{Case Study with Visualization}
To explicitly see how GRACE captures network clusters through the end-to-end combination of deep embedding, influence propagation, and self-training clustering models, we visualize the embedding spaces computed by GRACE at different steps on one of the ego-networks from the Facebook dataset.

\begin{figure*}[h!]
\centering
\subfigure[Original contents]{
\includegraphics[width=0.3\textwidth]{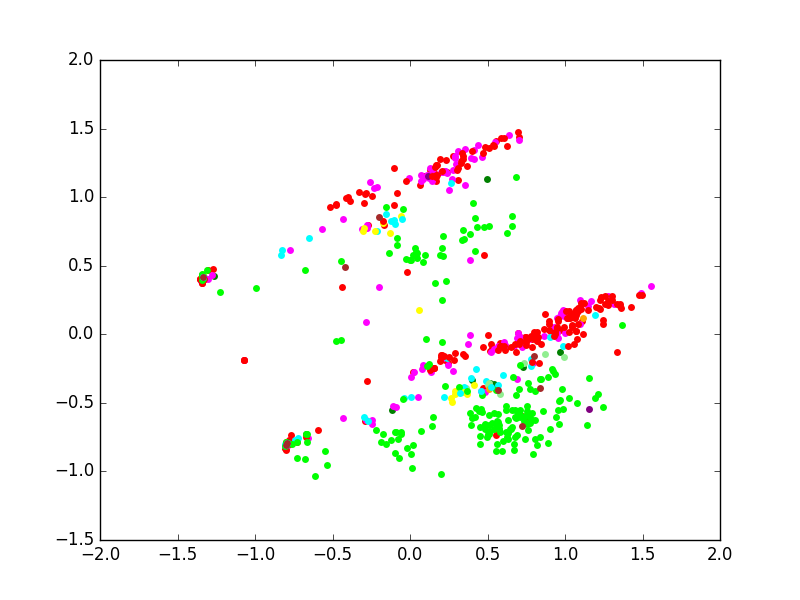}}
\subfigure[Jointly trained embedding]{
\includegraphics[width=0.3\textwidth]{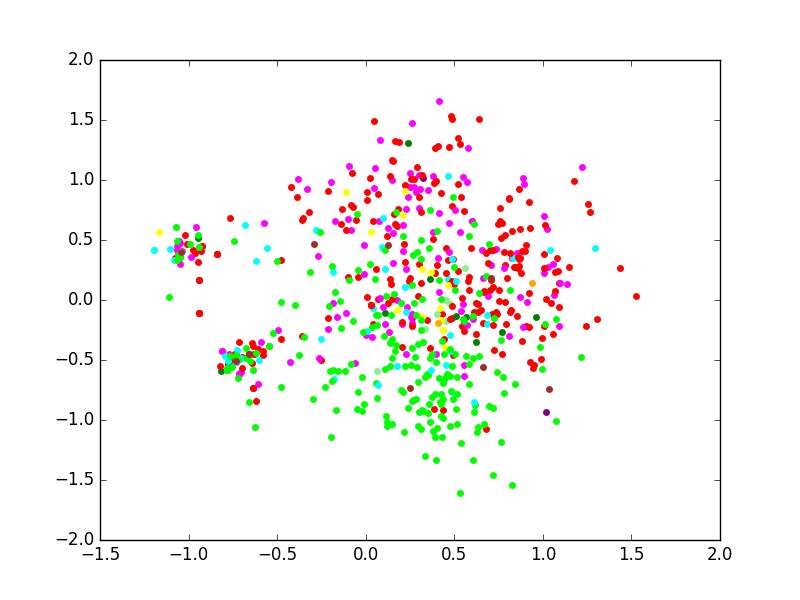}}
\subfigure[Propagated embedding]{
\includegraphics[width=0.3\textwidth]{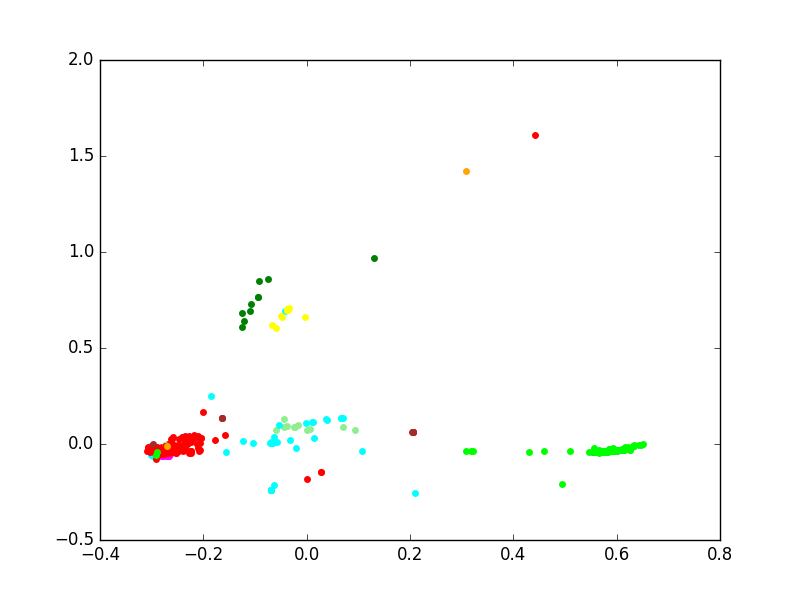}}
\caption{Visualization of embedding spaces computed on ego-network 1684 from the Facebook dataset.}
\label{fig:exp-vis}
\end{figure*}

Figure \ref{fig:exp-vis} visualizes the embedding spaces reduced to 2-dimensional through PCA \cite{jolliffe1986principal}. We use different colors and markers to draw nodes in different ground-truth clusters. As we can see, nodes from different clusters clutter a lot in the original content space. The situations become better after pre-training and co-training of the embedding and clustering models and reach the best after applying the influence propagation model, where nodes within the same ground-truth clusters distribute tightly in the embedding space, and those within different clusters lie far apart.
\section{Related Work}
We stress the novelty of this work and compare it with existing literature on two related lines of research.

\subsection{Network Community Detection}
Most traditional network community detection algorithms are unsupervised, based on either of the link density or content homogeneity assumptions, or both. The former assumes that communities should be constructed by densely connected nodes, and the latter requires that nodes in communities should also share similar contents.

Among various algorithms, the most widely used are based on heuristic metrics such as modularity \cite{girvan2002community, clauset2004finding} and maximal clique \cite{du2007community}.  They are fundamentally leveraging the link density assumption. As for content homogeneity, some algorithms firstly augment the graph with content links and then do clustering on the augmented graph \cite{ruan2013efficient, yang2009combining}, while some others attempt to find a partition that yields the minimum encoding cost based on information theory \cite{akoglu2012pics, rosvall2008maps}. \cite{liu2015community} leverages the concept of content propagation on networks to detect communities by firstly propagating the contents among neighboring nodes. The main drawback of these algorithms mainly lies in their high computational cost, which hinders the scaling to large networks with high-dimensional contents.

Another group of algorithms leverage probabilistic generative models. 
For instance, COCOMP \cite{zhou2012community} extends LDA \cite{blei2003latent} by taking community assignments as latent variables. It leverages content homogeneity by assuming that each community is a group of people that use a specific distribution of words. BigClam \cite{yang2013overlapping} does not make use of node contents at all. It leverages link density by modeling the assignments of communities solely based on existing links. A variation of BigClam called CESNA \cite{yang2013community} is later proposed, with additional consideration of simple binary-valued node attributes. It leverages both link density and content homogeneity by assuming communities to be formed by densely connected homogeneous nodes. 
Another method from the same group, Circles \cite{mcauley2012learning}, also leverages the same assumptions with a slightly different model. These shallow models are efficient even with large networks, but are still incapable of working with sparse noisy contents.

As a particular treatment to high-dimensional sparse contents with noise in large networks, we leverage the power of deep embedding. Based on our interpretation of networks as a consequence of influence propagation, the embedding model is able to efficiently explore node contents under the consideration of link structures and capture the intrinsic compact network representations oriented for high-performance graph clustering.


\subsection{Network Node Embedding}
The objective of node embedding is to find a compact node representation that captures the proximities among nodes on the networks. Traditional methods usually involve eigen-decomposition of the graph Laplacian matrix \cite{belkin2002laplacian, roweis2000nonlinear, tenenbaum2000global}. They have sound theoretical guarantees but do not scale well to real-world large networks. 

Recently, there has been a trend of neural network based graph embedding, started by the pioneering work of Deepwalk \cite{perozzi2014deepwalk}, which applies random walks on networks to capture node proximity and adopts Skip-gram \cite{mikolov2013distributed} from word embedding to compute node embedding. Its success motivates many subsequent studies which apply shallow neural networks like Skip-gram by either changing the random walk sampling methods \cite{grover2016node2vec, dong2017metapath2vec} or the proximity preserving objectives \cite{tang2015line, tang2015pte, cao2015grarep, yang2015network}. These models successfully inherit the training efficiency of neural networks and are effective in capturing node proximity based on simple network structures. However, as they only employ shallow neural networks, they are unable to deal with complex network structures and node contents.

Another group of node embedding algorithms leverage deep neural networks such as CNN mainly from computer vision \cite{krizhevsky2012imagenet, le2011learning} and RNN mostly from natural language processing \cite{mikolov2010recurrent, cho2014properties}. 
For instance, to leverage the CNN model, \cite{niepert2016learning} uses graph labelling to select an ordering of nodes, upon which fixed-sized receptive fields can then be computed, \cite{henaff2015deep, bruna2013spectral} utilize spectral filtering to emulate receptive fields in the Fourier domain, and \cite{kipf2016semi, defferrard2016convolutional, scarselli2009graph} treat convolution as neighborhood matching in the spatial domain. On the other hand, RNN have been utilized to embed sequences of nodes such as information cascade paths \cite{wang2017topological, li2017deepcas, li2015gated}. These models are often deep and powerful in exploring high-dimensional contents as well as complex network structures, but they usually assume a supervised or semi-supervised learning scenario, and their performance largely depends on the amount of labeled data.

In this work, we aim at unsupervised network clustering. To this end, we leverage deep embedding neural networks to model sparse noisy node contents. Through jointly training network embedding with a self-trained clustering component, the models are learned in an end-to-end unsupervised fashion.

\section{Conclusion}
Recently, there has been a trend in applying deep learning algorithms for graph modeling. To the best of our knowledge, we are the first to leverage deep embedding with influence propagation for unsupervised graph clustering. Our dynamic embedding model is able to integrate sparse noisy node contents and link structures. Both of them are vital for improving graph clustering performance, and they are in turn enhanced by the unsupervised clustering model through the self-training iterations in a closed loop. The framework fully leverages the power of deep neural networks and runs efficiently on GPUs.

While GRACE is designed for unsupervised graph clustering, it provides a promising way of deeply exploring network contents and links. Consider mining the web with rich page contents and complex hyperlinks, it is interesting to leverage our GRACE model for various down-stream applications, including page categorization, content profiling, hyperlink prediction and so on.
\bibliographystyle{ACM-Reference-Format}
\bibliography{grace} 

\end{document}